\documentclass[aps,twocolumn,showpacs,showkeys,superscriptaddress]{revtex4-1}
\usepackage[T1]{fontenc}
\usepackage{amsfonts,amsmath,amssymb,amsbsy}
\usepackage{graphicx,color}
\usepackage{times}
\usepackage[colorlinks=true, linkcolor=blue, citecolor=red, urlcolor=magenta]{hyperref}
\let\mathbf=\boldsymbol

\begin{document}

\preprint{UFV-UFOP-KSU Ribeiro et al.}

\title{Realization of Rectangular Artificial Spin Ice and Direct Observation of High Energy Topology}

\author{I. R. B. Ribeiro}
\affiliation{Laboratory of Spintronics and Nanomagnetism ($LabSpiN$), Departamento de F\'{i}sica,
Universidade Federal de Vi\c{c}osa, 36570-000 - Vi\c{c}osa - Minas
Gerais - Brazil.}
\affiliation{Instituto Federal do Esp\'{i}rito Santo, Alegre, 36570-900, Esp\'{i}rito Santo, 29520-000, Brazil }

\author{F.S. Nascimento} \affiliation{Departamento de F\'{i}sica,
Universidade Federal de Ouro Preto, 35931-008 - João Monlevade - Minas Gerais - Brazil.}

\author{S.O. Ferreira} \affiliation{Departamento de F\'{i}sica,
Universidade Federal de Vi\c{c}osa, 36570-000 - Vi\c{c}osa - Minas
Gerais - Brazil.}

\author{W. A. Moura-Melo}
\affiliation{Departamento de F\'{i}sica, Universidade Federal de
Vi\c{c}osa, 36570-000 - Vi\c{c}osa - Minas Gerais - Brazil.}

\author{C. A. R. Costa}
\affiliation{National Nanotechnology Laboratory (LNNano), National Center for Energy and Materials (CNPEM), Campinas, S\~ao Paulo, Brazil 13083-970}

\author{J. Borme}
\affiliation{INL-International Iberian Nanotechnology Laboratory, 4715-330, Braga, Portugal}

\author{P. P. Freitas}
\affiliation{INL-International Iberian Nanotechnology Laboratory, 4715-330, Braga, Portugal}

\author{G.\ M.\  Wysin }
\affiliation{Department of Physics, Kansas State University,
Manhattan, KS 66506-2601}

\author{C.I.L. de Araujo}
\email{dearaujo@ufv.br} \affiliation{Laboratory of Spintronics and Nanomagnetism ($LabSpiN$), Departamento de F\'{i}sica,
Universidade Federal de Vi\c{c}osa, 36570-000 - Vi\c{c}osa - Minas
Gerais - Brazil.}

\author{A. R. Pereira}
\email{apereira@ufv.br.}
\affiliation{Departamento de F\'{i}sica, Universidade Federal de
Vi\c{c}osa, 36570-000 - Vi\c{c}osa - Minas Gerais - Brazil.}

\date{\today}

\begin{abstract}
{In this letter, we have constructed and experimentally investigated
frustrated arrays of dipoles forming two-dimensional
artificial spin ices with different lattice parameters (rectangular
arrays with horizontal  and vertical lattice spacings denoted by $a$
and $b$ respectively). Arrays  with three different ratios $\gamma =a/b = \sqrt{2}$,
$\sqrt{3}$ and $\sqrt{4}$ are studied. Theoretical calculations of low-energy
demagnetized configurations for these same parameters are also presented.
Experimental data for demagnetized samples confirm most of the theoretical
results. However, the highest energy topology (doubly-charged monopoles) does not emerge
in our theoretical model, while they are seen in experiments for
large enough $\gamma$. Our results also insinuate that magnetic monopoles may be
almost free in rectangular lattices with a critical
ratio $\gamma = \gamma_{c} = \sqrt{3}$, supporting previous theoretical predictions.}
\end{abstract}

 \maketitle

Recently, the study of materials with frustrated
interactions has received a lot of attention in an attempt to understand new states of
matter\cite{Harris1997,Castelnovo2008,Ramirez99,Balents10,Wang2006,Mol2009,Mengotti2011,Moller2006}.
The main problem concerning the experimental investigation of the
properties of these structures is to find natural materials (in two
and three dimensions), which not only clearly exhibit frustration
but also provide reproducible results and adequate control for measurements.
It is not such a simple task. An alternative path was provided by techniques of nanotechnology,
in which artificial materials can be built with desirable properties and attributes in
order to permit the materialization of a large variety of different
sorts of geometrical frustration\cite{Wang2016,Nisoli2017}. Especially, artificial spin ices
in several different lattice geometries are important examples
\cite{Wang2006,Mengotti2011,Morgan2011}. They are two-dimensional
($2d$) arrays of elongated magnetic nanoislands, each containing an
effective magnetic moment or spin (see Fig.1) that mimics natural
three-dimensional ($3d$) spin ice materials\cite{Harris1997,Castelnovo2008,Ramirez99}.
\begin{figure}[h!]
    \centering
    \includegraphics[width=8.0 cm]{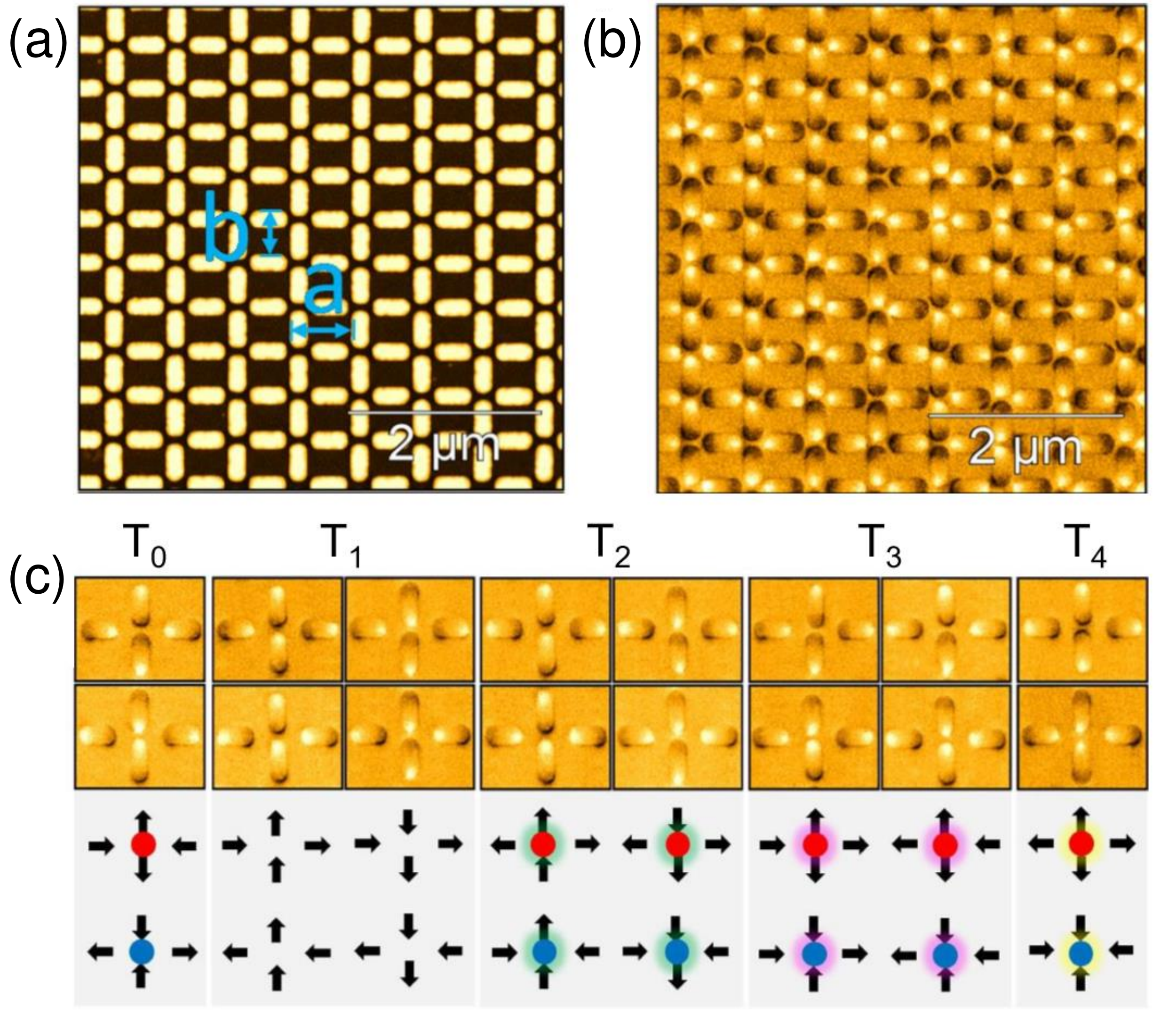}
    \caption{Artificial spin ice in a rectangular lattice. Consistent with other types of geometry
    (square, kagome etc), the ground state of a rectangular spin ice obeys the ice rule in all vertices,
    which dictates that two spins must point-in and the other two must point-out. Excited states violate of the ice rule. The particular array shown here has the aspect ratio $\gamma=a/b=\sqrt{2}$, which means
    that the ground state should exhibit residual charges in all vertices. (a) Atomic force microscope topography of a typical sample for $\gamma=\sqrt{2}$. (b) Picture from the magnetic force microscope of single domain permalloy magnetic nanoislands ($300 nm \times 100 nm \times 20 nm$). Bright and dark ends of each elongated nanoisland indicate the opposite poles and give the direction of the magnetic moment of the islands and (c) The five possible topologies in this system. }
    \label{fig:rede}
\end{figure}
However, such an artificial system in a $2d$ square
lattice is not completely frustrated since the ice rule (in which
two-spins must point-in and the other two must point-out in each
vertex) is not degenerate (the two topologies that obey the ice rule
have different energies\cite{Wang2006,Mol2009}) and, therefore, the
ice regime is not stabilized. Despite this, as in natural
spin ices, artificial square ice (and even other kinds of
artificial lattices) also supports quasiparticle excitations that are similar to
magnetic monopoles\cite{Mol2009,Morgan2011,Silva13,Mol2010,Nascimento12}, although, in
general, these monopoles are of different types in natural and
artificial materials. Indeed, $2d$ artificial square spin ice
supports excitations in which the oppositely charged monopoles occur
connected by observable and energetic strings (a kind of Nambu
monopole-antimonopole pair\cite{Mol2010,Silva12,Nambu74} in contrast to Dirac monopoles,
in which the string is not observable and does not have energy).
Therefore, it would be interesting to imagine and construct $2d$ artificial
lattices whose monopole pair excitations would have a string tension that tends
to vanish in such a way that, opposite magnetic charges would be
effectively interacting only by means of the usual Coulombic law.

A recent theoretical proposal was made to modify the square array into a
rectangular one \cite{Nascimento12}. Really,
such a deformation can tune the ratios of the interactions between
neighboring elements resulting in different magnetic ordering of the
system. Denoting the horizontal and vertical lattice spacings of the
rectangular array by $a$ and $b$ respectively, and defining a
parameter (the aspect ratio) that controls the stretching of the lattice $\gamma \equiv
a/b$, then, the ground state suffers a transition at
$\gamma=\sqrt{3}$ (or equivalently at $1/\sqrt{3}$ by interchanging $x$ and $y$ axes, or
make $\gamma\ge 1$ to avoid this ambiguity). The theoretical
calculations indicate that, for $1< \gamma< \sqrt{3}$, the ground
state (denoted $GSQ$) has residual magnetic charges (but not magnetic moments) in
all vertices, alternating from positive to negative in neighboring vertic
es.
Therefore, the total magnetic charge is zero. On
the other hand, for $\gamma > \sqrt{3}$, the ground state (denoted $GSM$) exhibits
alternating residual magnetic moments (but not charges) in all vertices and,
again, in this case, the total magnetic moment is zero. Exactly at the critical
value $\gamma= \gamma_{c}= \sqrt{3}$, the two different configurations $GSQ$ and $GSM$ have the
same energy and, therefore, the ground state at this particular $\gamma_{c}$ becomes degenerate,
suggesting a residual entropy at absolute zero temperature similar
to what happens in natural\cite{Harris1997,Castelnovo2008} and
($3d$) artificial\cite{Moller2006,Mol2010,Moler09} spin ice
materials. As a consequence, at $\gamma_{c}= \sqrt{3}$ (or $1/\sqrt{3}$),
the string tension connecting opposite magnetic charges tends to vanish
and, in principle, the monopoles become free to move. This
transition of the ground state is a consequence of the fact that,
differently from the square lattice (which has four distinct
topologies\cite{Wang2006} for the four spins meeting at each
vertex), the rectangular lattice exhibits five
topologies\cite{Nascimento12}: $T_{0},T_{1},T_{2},T_{3},T_{4}$ (see
Fig. 1c). The first two ($T_{0}$ and $T_{1}$) obey
the ice rule (two-in, two-out) with their energies depending on the
parameter $\gamma$. For $1<\gamma<\sqrt{3}$, the energy of $T_{0}$ is
smaller than the energy of $T_{1}$, while the contrary is valid for
$\gamma>\sqrt{3}$. In this letter we propose to realize an
experimental study of the ground state and excited states of rectangular lattices with different
ratios $\gamma$. Basically, we compare arrays with ratios $\gamma
<\sqrt{3}$ and $\gamma >\sqrt{3}$ to the array having the critical
value $\gamma=\gamma_{c}=\sqrt{3}$ (from now, dubbed $\gamma_{c}$-array).
For this comparison, we choose systems with lattices parameters having
ratios equal to $\gamma=\sqrt{2}$ and $\gamma=\sqrt{4}=2$.

For the fabrication of Permalloy nanoislands, a multilayer with
composition $Si$ / $Ta$ $3 nm$ / $Ni_{80} Fe_{20}$ $20 nm$ / $Ta$
$3nm$ was previously prepared by sputtering from tantalum (seed and
cap layer) and alloyed permalloy target, on silicon oxide substrate.
Then, the samples were covered with a $85 nm$ layer of $AR-N
7520.18$ negative tone photoresist and pattered by electron
lithography at $100 kV$ of acceleration voltage. After development,
the samples were etched by ion milling at 20$^\circ$ from normal
incidence, using secondary ion mass spectroscopy to detect the end
of the process. An ashing in oxygen plasma was subsequently
performed to remove the photoresist. The nanoislands dimensions of
$l = 300 nm$ and $w = 100 nm$ leads to saturation magnetization $780
\times 10^{3} Am^{-1}$, giving a magnetic moment $\mu = 4.68 \times
10^{-16} Am^{2} $ per island. Then, for the y-axis lattice spacing
$b=450 nm$ in our samples, the energy scale is $D =
\mu_{0}\mu^{2}/4\pi b^{3}= 2.4 \times 10^{-19}J$. The x-axis lattice
constant $a$ ranged from $636-900 nm$ in such a way that we have
investigated by magnetic force microscopy ($MFM$), $RASI$
arrangements with aspect ratios $a/b=\sqrt{2}, \sqrt{3}$ and
$\sqrt{4}$ (see Fig. 2). These systems were built on a area of $4
mm^{2}$ and the $MFM$ measurements performed in $25$ and $100 \mu
m^{2}$ area, which enabled topologies density analysis in arrays of
up $12 \times 22$ unit cell ($528$ islands). To improve the
statistics, the $MFM$ measurements were carried out in four
different regions of the samples. We have also done some Monte Carlo
numerical calculations of low energy configurations to compare with
experimental results.

\begin{figure}[h!]
    \centering
    \includegraphics[width=8.0 cm]{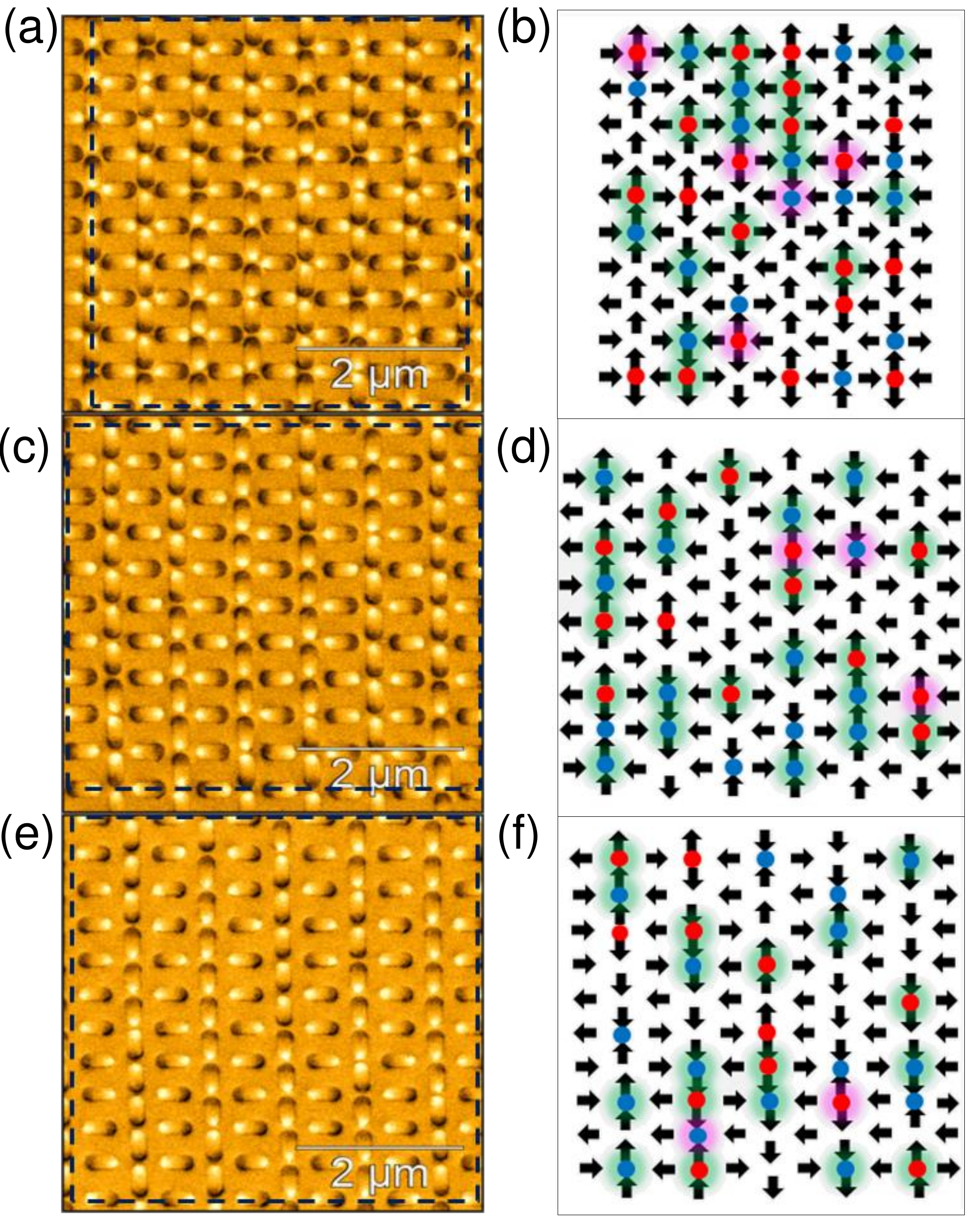}
    \caption{$MFM$ results of artificial spin ice in a rectangular lattice and representations of magnetic charges observed in each vertex with: a) and b) $\gamma=\sqrt{2}$; c) and d) $\gamma=\sqrt{3}$; e) and f) $\gamma=\sqrt{4}$.}
    \label{fig:rede}
\end{figure}

 To find low energy configurations of the nanoisland dipoles, an experimental demagnetization protocol
 was carried out with a commercial demagnetizer. In this process, the magnetic field is switched from
 positive to negative magnetic fields in sample plane at a frequency of $60Hz$, as the samples are
 moved away from the coil center. We meant to move the samples in a direction parallel to $a$, however
 due to the microstructure size of the samples, some misalignment can be expected. To optimize this
 procedure, we have tested two different demagnetization protocols\cite{NisoliRev13}. In the first,
 the sample is subjected to a sinusoidal magnetic field modulated by an exponential
 decay $h(t) = H_{max} \exp (-t) \cos (2\pi 60t)$, where $H_{max}$ represents the field to saturate
 the sample. In the second, the magnetic field strength was stepped down $(H_{max} - 0)$ in magnitude
 and switched in polarity with each step. However no substantial difference was found between the two
 protocols; so we adopted the first one to perform the experiments.

In the simulations we have considered each magnetic nanoisland as a macro Ising spin. These spins
interact via dipolar interactions. To obtain the evolution of the Ising spins under an external
magnetic field, we have adopted the same procedure employed by Budrikis \textit{et al.}\cite{Budrikis2012}.
In this consideration, one spin $\vec{S}^{i}$ can be flipped if the total field acting on it
satisfies $(\vec{h}_{ext}+ \vec{h}^{i}_{dip})\cdot \hat{S}^{i} < - h^{i}_{c} $, where $\hat{S}^{i}$
represents a unit vector along the spin direction, $\vec{h}_{ext}$ is the external
field, $\vec{h}^{i}_{dip}$ is the dipolar magnetic field produced by all spins of the lattice
at the position where spin $i$ is placed and $h^{i}_{c}$ is the island's switching barrier.
A perfect system is represented by a constant barrier while disorder can be implemented by
taking $h^{i}_{c}$ from a Gaussian distribution with standard deviation $\sigma$. Here we
consider disorder in the system to be absorbed into the dispersion of the switching barrier.

Using the samples, we analyzed the distribution of topologies and
the total magnetization for three demagnetized $RASI$ arrays studied
here ($a/b=\sqrt{2}, \sqrt{3}$ and $\sqrt{4}$). To accomplish that,
we computationally mapped the dipole configurations imaged by $MFM$
and assigned a value $m_{x} = \pm 1$ or $m_{y} = \pm 1$ to each
island moment, depending on the island magnetic orientation, as
presented in Fig.3a. The Table I summarizes the averaged
experimental results, obtained after analysis. There is a very low
total magnetization (in a range $0.03 - 0.10$, close to zero),
indicating a rather efficient demagnetization. Additionally, the
experimental data for the topology densities are very different from
those expected for arrays with randomly oriented individual moments
($n(T_{0}) = n(T_{4} = 12.5 \% $)) and ($n(T_{1}) = n(T_{2}) =
n(T_{3}) = 25 \% $); this also indicates that the demagnetization
protocol was successfully applied on the samples. Curiously, a few
number of $T_{4}$ topologies emerge for large enough $\gamma$, i.e.,
for $\sqrt{3}$ and $\sqrt{4}$ $RASI$. It is not seen for
$\gamma=\sqrt{2}$. The direct experimental observation of this
topology has never been predicted by our Monte Carlo calculations
presented below. In terms of real lattices and nanoislands, one
possible reason to explain the appearance of $T_{4}$ topology in
experiments (but not in simulations) is the significant reduction of
the energy scale between higher and lower energy of topologies
(Fig.1c).

For the ground state topologies ($T_{0}$ and $T_{1}$), we found that
$T_{1}$ topology as a function of $\gamma$ has a minimum at
$\gamma=\gamma_{c}=\sqrt{3}$. The same can be said for the density
of the $T_{0}$ topology (but with values about four times smaller
than the $T_{1}$ topology). On the other hand, by taking into
account the presence of monopole-antimonopole pairs in these systems
(excitations above the ground state associated with $T_{2}$ and
$T_{3}$ topologies), we notice that the pair density (the sum of
$T_{2}$ and $T_{3}$ densities) is greater for rectangular lattices
with the critical aspect ratio ($\gamma=\gamma_{c})$ than that
observed for others values of $\gamma$. Table I summarizes these
results, also indicating that critical $\gamma_{c}$-arrays exhibit
the maximum number of monopoles possible (i.e., the pair density and
the ground state topologies as a function of $\gamma$ would present
a peak and a minimum, respectively, at $\gamma=\gamma_{c}$).
Therefore, as expected, minimum values of the ground state
topologies are correlated with a maximum presence of excited states
(monopole topologies).

The experimental observations were taken at room temperature
(however, it is not important since these permalloy arrangements are
expected to be athermal). This suggests that the different numbers
of monopole pairs observed for different values of $\gamma$ results
from a purely geometrical effect, reinforcing the fact that
monopoles could be more spontaneously generated in
$\gamma_{c}$-arrays. Considering that the total energy of a pair
depends also on the energy of the string connecting the monopole
with its antimonopole, then, a reasonable hypothesis for this
geometrical influence on monopole number is that the string energy
decreases as $\gamma \rightarrow \gamma_{c}$, corroborating previous
theoretical results\cite{Nascimento12}, which predict very low
string tension for $\gamma_{c}$-arrays. Indeed, in Fig. 3b, in a
large section of a sample with $\gamma = \gamma_{c}$, one can
observe a great quantity of monopole-antimonopole pairs (most of
them with sizes $a$ or $b$) and a small quantity of isolated
monopoles.
\begin{figure}[h!]
    \centering
    \includegraphics[width=7.5 cm]{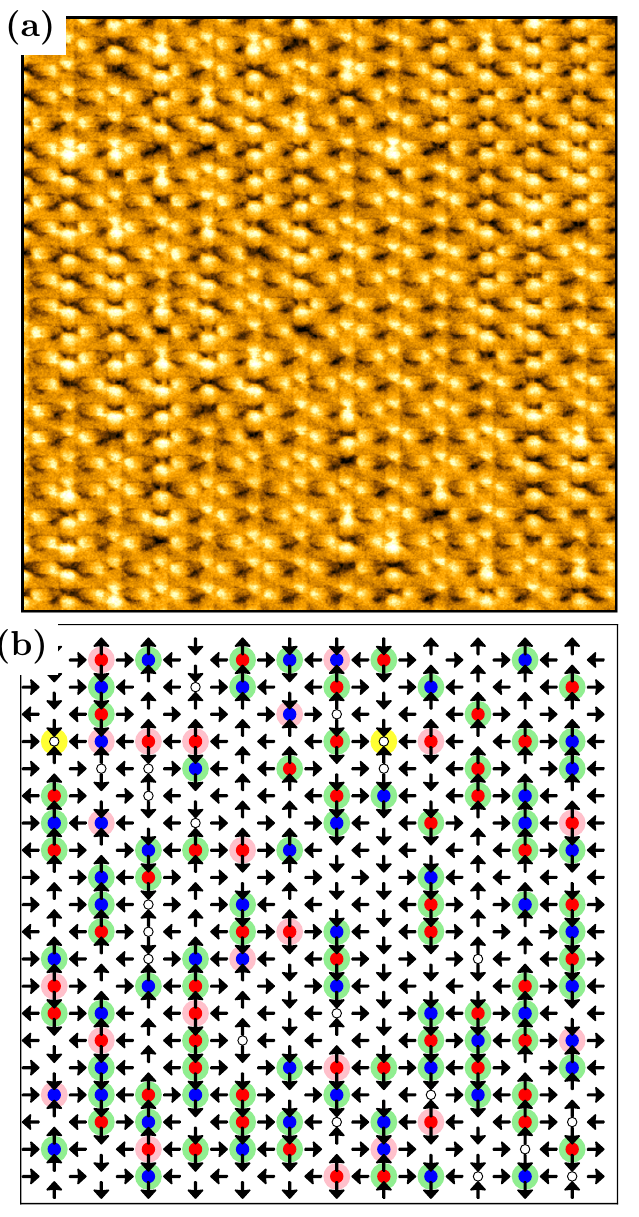}
    \caption{(a) Large area ($100 \mu m^{2}$) $MFM$ view of a $\gamma=\gamma_{c}=\sqrt{3}$
    sample and (b) magnetic moments and the topologies in each vertex, mapped computationally.}
    \label{fig:rede}
\end{figure}
\begin{table}[t]
\centering
\caption{\label{TabMagTop} Summary of the experimental results for magnetization and
topologies density for $a/b = \sqrt{2}, \sqrt{3}$ and $\sqrt{4}$.}
\vspace{0.5cm}
\begin{tabular}{l|ccccc}
    \hline
    $a/b$ & T0 & T1 & T2 & T3 & T4 \\
    \hline
    $\sqrt{2}$ & 0.16911 & 0.45456 & 0.25472 & 0.11731 & 0.00430 \\
    $\sqrt{3}$ & 0.09050 & 0.40643 & 0.38662 & 0.10651 & 0.00994 \\
    $\sqrt{4}$ & 0.12656 & 0.43257 & 0.35819 & 0.08020 & 0.00248 \\
    \hline
\end{tabular}
\end{table}
We also carried out Monte Carlo ($MC$) calculations using macro Ising spins for the island dipoles
to compare with experiments. To be closer to the experimental situation described above, a demagnetization
field is included in the simulations. This differs from the earlier calculations\cite{Nascimento12},
which do not consider external fields. Figure 4 shows the topology densities after having applied the
demagnetization procedure in the $MC$ simulations and its comparison with the topology densities
measured by $MFM$. We can see that the final topologies depend significantly on $\alpha$, which
is the angle that the external magnetic field is applied in relation to the larger lattice
spacing (horizontal or $a$-axis in Fig.2). For $\alpha < 0.15 \pi$ ($\alpha > 0.35 \pi$) the
energetic flow occurs only on horizontal (vertical) dipoles. Of course, such behavior is a
consequence of the fact that, if the external field is too oblique in relation to the horizontal
dipoles, the projection of this field along the perpendicular dipoles will not be sufficient to
overcome the islands' switching barriers $h_{c}^{i}$, so the perpendicular dipoles will be frozen,
i.e., they will maintain the initial configuration. We should notice that the $MC$ simulations do
not include the effects of thermal fluctuations, which might explain why the $T_{4}$ topology is
not reproduced by them. Perhaps even minor thermally induced fluctuations would be enough to help
to produce the doubly-charged poles.

Initially, samples are magnetized in a diagonal direction, which
implies that the topology densities are initialized to the values
$n(T_{1}) = 1$ and $n(T_{0}) =  n(T_ {2}) = n(T_{3}) = n(T_{4}) =
0$. However, for $0.25\pi < \alpha < 0.35\pi$, the energetic flow is
distributed for all the system. In Fig.4a we note that the
experimental data are more similar to the theoretical results for
$\alpha\approx 0.2\pi$, although, even for this case, our
simulations were not able to exhibit the $T_{4}$ topology and yet,
the density of the $T_{3}$ topology is also very small, arising only
for large enough $\alpha$. Therefore, the highest topologies are
responsible for the biggest contrast between theory and experiments.
Maybe the system sizes used in our calculations are too small to get
good statistics for the topologies of low probabilities. Figure 4b
shows the experimental and theoretical behavior of topology
densities for a range of lattice spacings and fixed
$\alpha=0.20\pi$. In overall, the theoretical results for the ground
state topologies ($T_{0}$ and $T_{1}$) are in good qualitative and
quantitative agreement with experimental data. However,
theoretically, the $T_{0}$ density goes slowly from approximately
$0.20$ for $\gamma=\sqrt{2}$ to almost zero (for $\sqrt{4}$), while
experimentally (Table I and Fig.4b), this density varies from $0.16$
for $\gamma=\sqrt{2}$, decreasing to $0.09$ for $\gamma =\sqrt{3}$
(similar to theoretical results) but, it turns to increase again to
$0.12$ for $\gamma = \sqrt{4}$. Therefore, there is an important
qualitative difference between our simulations and experiments in
the region $\gamma > \sqrt{3}$. For the density of the $T_{1}$
topology (green line), the $MC$ simulations indicate that it becomes
practically constant (around $0.60$) as $\gamma$ varies, while
experimental data (see again the Table I) remains almost constant
with $[n(T_{1})]$ varying near above $0.4$. Furthermore, considering
the monopole excitations ($T_{2}$ and $T_{3}$ topologies), we
observe a good quantitative agreement between the $MC$ simulations
and experiments (Fig.4b and Table I) only for $T_{2}$-type monopole
(red line). For $T_{3}$ topology (cyan line), the simulations lead
to a very low density as compared to experiments. Despite the
differences pointed out here, we can say that, in general, there is
an overall qualitative (and even quantitative) agreement between the
simple Ising spin model for magnetic nanoislands used here and our
experimental data. These agreements become better in the region $1 <
\gamma < \sqrt{3}$. Finally, we have also calculated the energy of
the topologies as a function of $\gamma$ (see Fig.4c). They indicate
that, independently of $\gamma$, the energy for creating $T_{3}$
monopoles is bigger than the energy for creating $T_{2}$ monopoles.
It may explain the lower presence of $T_{3}$ excitations around the
lattice in both theoretical and experimental results. In addition,
the energy of doubly-charged monopoles ($T_{4}$ topology) is the
biggest one, but it decreases relatively rapidly as $\gamma$
increases. Such behavior, to some extent, justifies the direct
observation of these $T_{4}$ excitations in experiments for $\gamma$
large enough ($\gamma=\sqrt{3}$ and $\sqrt{4}$, see Table I).
\begin{figure}[!h]
    \centering
    \includegraphics[width=10 cm]{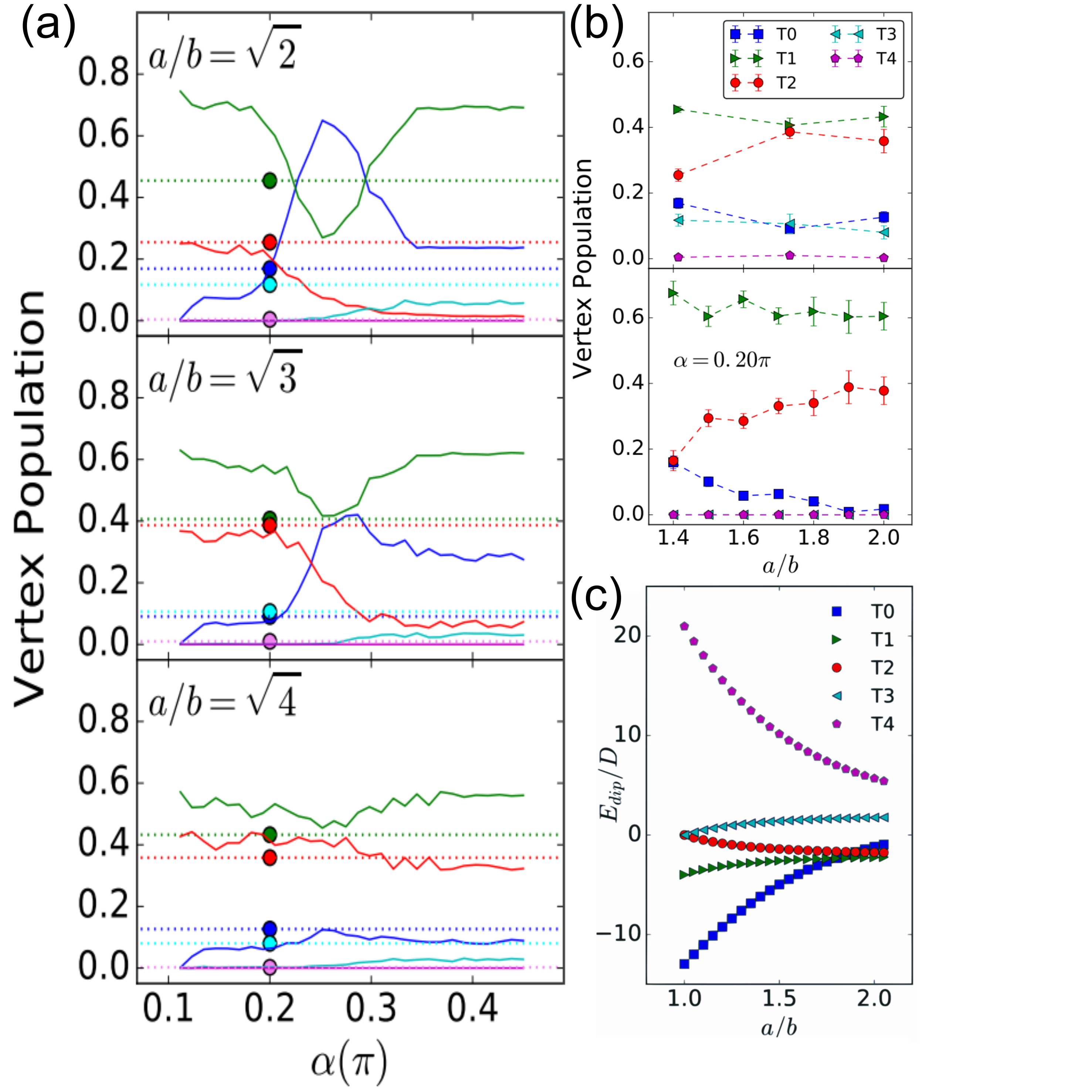}
    \caption {(a) Vertex population densities as functions of angle $\alpha$ between the
    demagnetizing field and the unitary cell along the $x$-axis for:
    (Top to bottom), $\gamma = \sqrt{2}$,  $\gamma = \sqrt{3}$ and  $\gamma =
    \sqrt{4}$. The colored circles represent the experimental data
    for $\alpha = 0.2 \pi$.(b) Vertex population densities as functions of $\gamma$ for the
    demagnetizing field fixed at $\alpha=0.2\pi$, for experimental (top) and theoretical (bottom) results.
    (c) Topology energies as functions of $\gamma$.}
    \label{fig:rede}
\end{figure}

In summary we have experimentally and theoretically investigated
two-dimensional artificial spin ices in rectangular lattices. Due to
the possible misalignments during the experimental demagnetization
protocol, we have performed our theoretical demagnetization scheme
by using an external magnetic field switched along a direction at
different angles $\alpha$ in relation to the larger cell side. The
topology densities of the experimental samples (numerically counted
from the $MFM$ measurements realized in samples with $a/b =
\sqrt{2}, \sqrt{3}, \sqrt{4}$) were compared to the topology
densities obtained theoretically by deadened sinusoidal external
fields. The overall qualitative agreement between the simple
theoretical model and experimental results is remarkable. A
quantitative agreement is better achieved mainly when the magnetic
field is applied at an angle close to $\alpha = 0.20\pi$ in relation
to the larger cell side. Therefore, in general, we can say that the
experimental results confirm the simple theory of Ising spin islands
most used nowadays, but interestingly, topology $T_{4}$
(doubly-charged monopole), which has the highest energy, could be
seen only in experiments for lattices with large enough $\gamma$.
Concerning this fact, $MC$ simulations are able to give, at least, a
route for this experimental visualization, showing that the energy
of the $T_{4}$ topology decreases considerably as $\gamma$
increases. Of course, some disagreements between the theory
developed here and experiments are to be expected in view of the
exceedingly complex samples as compared to the simple theoretical
approach. Our results also provide experimental evidence that
magnetic monopoles may be almost free in rectangular lattices at the
critical aspect ratio. Indeed, the density of magnetic monopoles
(topologies $T_{2}$ and $T_{3}$) is purely a geometrical effect,
having a maximum at an intermediate array ($\gamma_{c}$-array). Such
a phenomenon may be associated with the fact that the string tension
tends to vanish as $\gamma \rightarrow \gamma_{c}$, lending support
to previous theoretical predictions\cite{Nascimento12}.


\section*{Acknowledgments}

The authors would like to thank the Brazilian agencies CNPq, FAPEMIG
and CAPES.


\begin{thebibliography}{99}



\bibitem{Harris1997} M. Harris, S. Bramwell, D. McMorrow, T. Zeiske, and K. Godfrey, Phys. Rev. Lett. 79, 2554 (1997).

\bibitem{Castelnovo2008} C. Castelnovo, R. Moessner, and L. Sondhi, Nature 451, 42 (2008).

\bibitem{Ramirez99} A.P. Ramirez, A. Hayashi, R.J. Cava, R. Siddharthan, and B.S. Shastry. Nature, 399:333 (1999).

\bibitem{Balents10} L. Balents. Nature, 464:199 (2010).

\bibitem{Wang2006} R. F. Wang, C. Nisoli, R. S. Freitas, J. Li, W. McConville, B. J. Cooley, M. S. Lund, N. Samarth, C. Leighton, V. H. Crespi, and P. Schiffer, Nature \textbf{439}, 303 (2006).

\bibitem{Mol2009} L.A. M\'{o}l, R.L. Silva, R.C. Silva, A.R. Pereira, W.A. Moura-Melo, and B.V. Costa, J. Appl. Phys. \textbf{106}, 063913 (2009).

\bibitem{Mengotti2011} E. Mengotti, L.J. Heyderman, A.F. Rodriguez, F. Nolting, R.V. Hugli, and H.B. Braun, Nat. Phys. \textbf{7}, 68 (2011).

\bibitem{Moller2006} G. M\"{o}ller and R. Moessner, Phys. Rev. Lett. \textbf{96}, 237202  (2006).

\bibitem{Wang2016} Y.-L. Wang, Z.-L. Xiao, A. Snezhko, J. Xu, L. E. Ocola, R. Divan, J. E. Pearson, G. W. Crabtree, and W.-K. Kwok, Science \textbf{352}, 962 (2016).


\bibitem{Nisoli2017} C. Nisoli, V. Kapaklis, and P. Schiffer, Nat. Phys. \textbf{13}, 200 (2017).

\bibitem{Morgan2011} J.P. Morgan, A. Stein, S. Langridge, and C. Marrows, Nature Phys. \textbf{7}, 75 (2011).

\bibitem{Silva13} R. C. Silva, R. J. C. Lopes, L. A. S. L.A. M\'{o}l , W. A. Moura-Melo, G. M. Wysin, and A. R. Pereira, Phys. Rev. B \textbf{87},

\bibitem{Mol2010} L.A.S. M\'{o}l, W.A. Moura-Melo, and A.R. Pereira, Phys. Rev. B \textbf{82}, 054434 (2010).

\bibitem{Nascimento12} F.S. Nascimento, L.A.S. M\'{o}l, W.A. Moura-Melo, and A.R. Pereira, New J. Phys.  \textbf{14},115019 (2012).

\bibitem{Silva12} R.C. Silva, F.S. Nascimento, L.A. S. M\'{o}l, W.A. Moura-Melo, and A.R. Pereira, New J. Phys. \textbf{14}, 015008 (2012).

\bibitem{Nambu74} Y. Nambu, Phys. Rev. D 10, 4262 (1974).

\bibitem{Moler09} G. M\"{o}ller and R. Moessner, Phys. Rev. B \textbf{80}, 140409(R) (2009).

\bibitem{NisoliRev13} C. Nisoli, R. Moessner, and P. Schiffer, Rev. Mod. Phys. \textbf{85}, 1473 (2013).

\bibitem{Budrikis2012} Z. Budrikis, K.L. Livesey, J.P. Morgan, J. Akerman, A. Stein, S. Langridge, C.H. Marrows and R.L. Stamps, New J. Phys.  \textbf{14}, 035014 (2012).




\end{thebibliography}
\end{document}